\documentclass[preprint,
superscriptaddress,
 amsmath,amssymb,
 aps,
 pra
]{revtex4-1}
\usepackage{graphicx}
\usepackage{xcolor}
\usepackage{dcolumn}
\usepackage{bm}
\usepackage{amsmath}
\usepackage{graphicx}
\usepackage{bbold}
\usepackage{float}
\usepackage{physics}
\usepackage[colorlinks=true, allcolors=blue]{hyperref}

\begin{document}

\title{Phase diagram and elementary excitations of strongly interacting droplets with non-local interactions}

\author{Maciej Łebek}
\email{maciej.lebek@fuw.edu.pl}
\affiliation{Faculty of Physics, University of Warsaw, Pasteura 5, 02-093 Warsaw, Poland}
\affiliation{Center for Theoretical Physics, Polish Academy of Sciences, Al. Lotnik\'{o}w 32/46, 02-668 Warsaw, Poland}

\author{Jakub Kopyciński}
\affiliation{Center for Theoretical Physics, Polish Academy of Sciences, Al. Lotnik\'{o}w 32/46, 02-668 Warsaw, Poland}

\author{Wojciech Górecki}
\affiliation{Faculty of Physics, University of Warsaw, Pasteura 5, 02-093 Warsaw, Poland}

\author{Rafał Ołdziejewski}
\affiliation{Max Planck Institute of Quantum Optics, 85748 Garching, Germany}
\affiliation{Centre for Quantum Optical Technologies, Centre of New Technologies, University of Warsaw, S. Banacha 2c, 02-097 Warsaw, Poland}

\author{Krzysztof Pawłowski}
\affiliation{Center for Theoretical Physics, Polish Academy of Sciences, Al. Lotnik\'{o}w 32/46, 02-668 Warsaw, Poland}

\date{\today}

\begin{abstract}
A one-dimensional bosonic gas with strong contact repulsion and attractive non-local interactions may form a quantum droplet with a flat-top density profile. We focus on a system in the Tonks-Girardeau limit of infinitely strong contact repulsion. We show that the main system features
are the same for a broad class of non-local interaction potentials. Then, we focus on a limiting case, the one of slowly varying density profiles, to find approximate formulas for the surface and bulk energies of a droplet. We further characterise the system by numerically finding the excitation spectrum. It consists of two families: phononic-like excitations inside droplets and scattering modes. Analysis within the linearised regime is supplemented with the full, nonlinear dynamics of small perturbations.
\end{abstract}
\maketitle

\tableofcontents
\section{Introduction }
Quantum droplets are a prime example that the mean-field (MF) description may fail even for weakly interacting Bose gas ~\cite{petrov_quantum_2015,lima_beyond_2012,lima_quantum_2011,bottcher_new_2021}. The MF approach predicts an unstable weakly interacting system, where the corresponding attractive and repulsive contributions nearly cancel each other. This gives way to the enhanced role of zero-point energy fluctuations~\cite{lima_beyond_2012,lima_quantum_2011,fischer}, usually named the Lee-Huang-Yang (LHY) term~\cite{lee_eigenvalues_1957}, stabilising the emerging droplet. Quantum liquids, however, are more robust in lower dimensions due to the increased quantum fluctuations~\cite{Petrov2016,Santos2017,Malomed2018,Zin2018,Ilg2018,Rakshit2019,Edmonds2020,morera_quantum_2020,morera_universal_2021,guijarro_ultradilute_2022,Ferlaino2022}. Particularly, in 1D, quantum droplets have been theoretically studied for strongly correlated systems with non-local interactions both in continuum and in the presence of an optical lattice~\cite{oldziejewski_strongly_2020,dePalo2020,de_palo_formation_2022,morera_superexchange_2022}. 

Historically, dipolar gases represent the first ultra-cold system with experimentally observed non-local interactions \cite{Griesmaier2005}. Subsequent advancements in cooling elements like erbium \cite{Ferlaino2012} and dysprosium \cite{Lev2011} have boosted the entire field, that now enters the era of Bose-Einstein condensates of polar molecules \cite{Bigagli2024Jul}. The latter system may become the platform to study droplets with strong non-local interactions \cite{Langen2024Jul}.
For the time being, the efforts of theorists are largely inspired by experiments on quasi-1D dipolar gases conducted in B. Lev's group \cite{Kao296, kuan-yu2023}, but
many-body quantum systems with non-local interactions can be realised in other physical scenarios. These include photon-mediated interactions in cavities \cite{Vaidya2018}, interactions in Rydberg atoms \cite{Low2012May}, systems in artificial dimensions \cite{Barbiero2020}, and strong non-local interactions between excitations in solids \cite{Eisenstein2004Dec}. Systems exhibiting non-local forces are employed in various implementations of the extended Bose-Hubbard model \cite{Chanda2024May}. 

There is a rich literature about a quasi-1D ultracold gas with dipolar interaction responsible for non-local potential. In these cases usually an effective dipolar interaction potential is employed~\cite{Deuretzbacher2010, Deuretzbacher2013erratum} \footnote{The proper analysis of the scattering of dipolar atoms in quasi 1D, and rigorous justification for using an effective dipolar potential is still missing.}. Different approaches, for instance, based on the Bogoliubov approach \cite{edler_quantum_2017}, or using the exact results from the Lieb-Liniger model partially neglecting quantum fluctuations \cite{oldziejewski2020, Kopycinski2022,Kopycinski2023prl} or
approximating them using correlation functions of non-dipolar systems 
 \cite{dePalo2020} lead to the same qualitative results. All papers predict the emergence of 1D droplets across all interaction regimes, being ultra-dilute self-bound objects (with negative energy for an untrapped state) marked by a flat-top density profile~\cite{bottcher_new_2021}. The emergence of such 1D droplets is quite a generic result of the competing interaction.
 
Our main goal is to show the common features of 1D droplets arising due to different non-local interaction potentials. We focus on the excitations of a droplet. Elementary excitations govern its low-energy dynamics, characterise the response to small perturbations and can be used to study the low-temperature thermodynamics. In 1D they play a crucial role in the distinction between  between a droplet and a bright soliton \cite{edler_quantum_2017,tylutki_collective_2020,oldziejewski_strongly_2020}. Both are bound states, present for an arbitrary number of particles in different interaction regimes. Namely, a droplet solution appears when short-range repulsion prevails over long-range attraction, contrary to the bright soliton case. Although quantum droplets in 1D exhibit a characteristic flat-top density profile for greater numbers of particles, for smaller systems, one can only distinguish between a bright soliton and a droplet by studying their excitation spectrum. 
This is motivated by the fact that in a one-dimensional Bose system, at least in the case without LHY corrections, there are no collective modes in a bright soliton -- there are continuum modes solely~\cite{Castin2001}. In contrast, 
the excitation spectrum of a droplet supports small-amplitude collective excitations~\cite{tylutki_collective_2020}.
Moreover, the excitation spectrum differs between 1D and 3D droplets, for the latter supports both bulk and surface modes \cite{Bulgac2002,petrov_quantum_2015}. The complementary problem of solitonic excitations in such systems was studied in~\cite{Kopycinski2022b,Edmonds2023, kopycinski2024qmc,Katsimiga2023,kopycinski2024prl}. 

Little is known, however, how the above picture changes by introducing strong correlations between the particles. 
In the case of non-dipolar system, with purely repulsive contact interaction, one can experimentally show that while tuning from the weakly to the strongly interacting regime via a magnetic Feshbach resonance, the dispersion changes dramatically by emerging hole-like excitations absent for higher dimension and weaker interactions~\cite{meinert2015}. Then for
dipolar system, the excitation spectra may change substantially even for weakly interacting gas, e.g. one observes the roton spectrum \cite{Petter2019, Chomaz2023}, leading to supersolid transition \cite{Guo2019Oct, Boettcher2019, bottcher_new_2021, Chomaz2019, Tanzi2021Mar}. It is however already known that the spectrum may substantially change in the strongly interacting case \cite{Oldziejewski2018}. 

In this work, we investigate a one-dimensional Bose gas with competing strong contact repulsion and arbitrary, relatively weak, non-local attractive interactions --  modelled by  Gaussian, exponential, and dipolar-like interaction potentials.  We discuss universal properties of the system to identify crucial physical parameters and indicate regimes where the system may be tractable analytically and numerically via simple approximate models. We focus on the latter regimes to discuss the appearance of liquid ground states and their properties. 
Insights from the ground state analysis motivate a very simple rectangular ansatz for the density profiles of droplets. Within approximation of such ansatz, we derive simple and useful formulas for the droplet width and energy. 
Then, we turn to the Bogoliubov-de Gennes (BdG) equations to analyze the excitation spectrum, obtaining expressions for the dispersion relations and the number of phononic bound modes that a droplet can exhibit. In addition to the study of excitations within the framework of BdG equations, we numerically analyse the response to the initial perturbation in the full, nonlinear dynamics.

The paper is organized as follows. In Sec.~\ref{sec:model} we introduce the most important features of the system under study and analyse phase diagram assuming a homogenous system. We briefly describe the non-local interactions in one dimension and the mean-field like description of strongly interacting Bose gas. This is followed by the analysis of the inhomogenous droplet ground state of the system in Sec.~\ref{sec:inhomogeneous}. In particular, we specify the regime, for which the density profile of the droplet can be found exactly and introduce the rectangular ansatz for the density profiles. After that, in Sec.~\ref{sec:exc} we present the results for the excitation spectrum obtained from the solution of BdG equations. In Sec.~\ref{sec:nonlin} we broaden this analysis by studying the response to initial perturbation in dynamics given by the nonlinear, hydrodynamic equation. Finally, in Sec.~\ref{sec:summ} we give a summary of results presented in the paper.

\section{The model and phase diagram}\label{sec:model}
We consider a system of $N$ bosonic atoms with mass $m$ confined to a one-dimensional box with periodic boundary conditions. The atoms interact via repulsive contact potential and attractive long-range forces, so the interaction potential reads

\begin{equation}
    V(x-x')=V_{\rm contact}(x-x')-V_\sigma(x-x'),
\end{equation}
where
\begin{equation}
    V_{\rm contact}(x-x')=g \delta(x-x'), \qquad   V_\sigma(x-x')=\frac{\lambda}{\sigma} \, \mathcal{V} \left(\frac{x-x'}{\sigma} \right), \qquad \lambda ,g, \sigma>0.
\end{equation}
The parameter $\sigma$ defines the characteristic range of the long-range interaction, and couplings $g, \lambda$ measure the strength of contact and long-range interactions, respectively. Moreover, we assume that potential $\mathcal{V}(x)$ is integrable and $\int {\rm d} x \,\mathcal{V}(x) =1$.

The model for $\lambda=0$ corresponds to the well-studied Lieb-Liniger (LL) model with coupling $g$, whose ground state properties are well understood. The Hamiltonian can be thus written as
\begin{equation}
    \hat{H}= \hat{H}_{\rm LL} - \frac{1}{2} \int {\rm d} x {\rm d}y V_\sigma(x-y) \hat{n}(x) \hat{n}(y),
\end{equation}
where $\hat{n}(x)$ are the standard density operators.

In our approach, we build on the knowledge from LL model and add weak, long-range interactions perturbatively. In the first order, this yields the modified ground state energy per particle, similarly to \cite{dePalo2020}:
\begin{equation}
    \varepsilon(\rho) :=\frac{E(\rho)}{N} = \frac{\langle \psi^{ \rm GS}_{ \rm LL}| \hat{H}| \psi^{\rm GS}_{\rm LL} \rangle}{N} = \frac{E_{\rm LL}(g,\rho)}{N} - \frac{\rho}{2} \int {\rm d}z  V_\sigma (z) g^{(2)}_{\rm LL}(g,\rho;z),
\end{equation}
where $E_{\rm LL}(g,\rho)$ is the ground state energy of the LL model at the density $\rho$ and $g^{(2)}_{\rm LL}(g,\rho;z)$ is the ground state two-particle correlation function. It depends only on the relative distance $z$ between the particles, due to translational invariance of the ground state. 
These quantities can be accurately approximated with known analytical functions ~\cite{Lang2017, ristivojevic_conjectures_2019}. Their exact, closed forms are known, for both energy and the correlation function, only for $g=0$, and for our main case of interest, which is $g \to \infty$. 
In such a regime, the LL gas can be mapped onto Tonks-Girardeau (TG) gas~\cite{Girardeau1960}. Energy and pair correlation function read
\begin{equation}
    \varepsilon_{\rm TG}( \rho) = \frac{\hbar^2 \pi^2}{6m} \rho^2,  \qquad g_{\rm TG}^{(2)}( \rho; z) = 1- \left(\frac{\sin(\rho \pi z)}{\pi \rho z} \right)^2,
\end{equation}
and thus ground state energy functional takes the form:
\begin{equation}
    \varepsilon(\rho) = \frac{\hbar^2 \pi^2}{6m} \rho^2 - \varepsilon_{\rm LR}(\rho), \qquad    \varepsilon_{\rm LR}(\rho) = \frac{\rho}{2} \int {\rm d}z  V_{\sigma}(z) g_{TG}^{(2)}(\rho;z).
    \label{eq:energy-functional-TG}
\end{equation}
Now, let us rewrite the long-range contribution by changing the variables as $t=z/\sigma$

\begin{equation}
    \varepsilon_{\rm LR}(\rho) = \frac{\rho \lambda}{2} \int {\rm d}t \mathcal{V}(t) \bigg[1- \left(\frac{\sin(\rho \pi \sigma t)}{\pi \rho \sigma t} \right)^2 \bigg] = \frac{\rho \lambda}{2} f_{\mathcal{V}}(\kappa),
\end{equation}
where $\kappa= \rho \sigma$ and 
\begin{equation}\label{eq:ffunction}
    f_\mathcal{V}(\kappa) = \int {\rm d}t \mathcal{V}(t) \bigg[1- \left(\frac{\sin(\pi \kappa t)}{\pi \kappa t} \right)^2 \bigg].
\end{equation}
The full energy functional thus reads
\begin{equation}
    \varepsilon(\rho) = \frac{\hbar^2 \pi^2}{6m} \rho^2 - \frac{\lambda}{2} \rho f_\mathcal{V}(\rho \sigma).
\end{equation}
This form tells us that all information about the potential is contained in the function $f_\mathcal{V}$. What is more, we find that the range of the potential $\sigma$ enters the problem only as a product with $\rho$. This shows that physically relevant parameter is the  ratio $\kappa$ between $\sigma$ and interparticle distance $\rho^{-1}$. It  has the interpretation of the number of atoms within the range of the non-local potential $\sigma$.

\begin{figure}[H]
    \centering
    \includegraphics[scale=0.89]{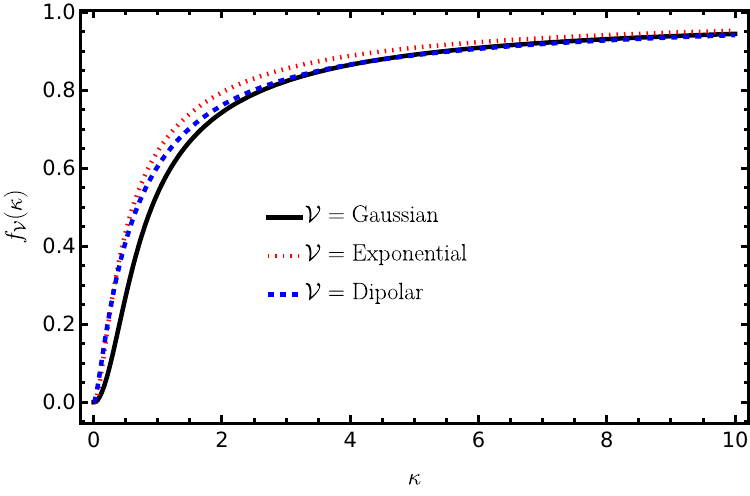}
    \caption{Function $f_\mathcal{V}(\kappa)$ for three potentials: Gaussian, Exponential and Dipolar. All functions tend to 1 at large arguments. Notably, the dipolar case is well approximated by the case of exponential potentials for small $\kappa$ and by Gaussian case for large $\kappa$.}
    \label{fig:1}
\end{figure}

The zero temperature phase diagram in the case $g\to\infty$ is solely determined by the energy functional Eq.~\eqref{eq:energy-functional-TG}. Before we proceed, let us note some universal features of function $f_\mathcal{V}$, that should hold for an arbitrary potential. We will illustrate the main properties of $f_\mathcal{V}$ with three exemplary interaction potentials (see Fig.~\ref{fig:1}):
\begin{itemize}
    \item \text{Gaussian} $\quad \mathcal{V}(x) =\frac{1}{\sqrt{\pi}} e^{-x^2}, \qquad f_\mathcal{V}(\kappa)=1- \frac{-1+e^{-\pi^2 \kappa^2}+ \pi^{3/2} \kappa \text{Erf}( \pi \kappa)}{ \pi^2 \kappa^2}$
    \item \text{Exponential} $\quad \mathcal{V}(x) =\frac{1}{2} e^{-x}, \qquad f_\mathcal{V}(\kappa)=1+\frac{-4\pi \kappa \arctan(2 \pi \kappa) +\log(1+4\pi^2 \kappa^2)}{4 \pi^2 \kappa^2}$
    \item \text{Dipolar} 
        $\label{eq:dip-potential}
        \quad \mathcal{V}(x) = \frac{1}{4}\bigg(-2|x|+\sqrt{2 \pi} (1+x^2)e^{x^2/2}\text{Erfc}(|x|/\sqrt{2}) \bigg), \qquad f_{\mathcal{V}}(\kappa) \quad \text{ numerically}.$
        
        This potential is widely used for ultracold dipolar atoms in a quasi-one-dimensional trap and is derived by integrating the three-dimensional dipole-dipole interaction over the $y$ and $z$ coordinates, assuming atoms occupy ground state Gaussian orbitals in these directions~\cite{deuretzbacher_ground-state_2010}.
\end{itemize}

For the first two cases $f_\mathcal{V}$ can be computed analytically, for the dipolar case one has to calculate it numerically. Qualitatively, the difference between functions $f_\mathcal{V}$ for these potentials is small, as it is visible from Fig.~\ref{fig:1}.
First of all, in the limit of large $\kappa \to \infty$ we have $f_{\mathcal{V}}(\kappa) \to 1$. This is because $\mathcal{V}(x)$ integrates to $1$, and for large $\kappa$ (i.e. $\sigma \gg \rho^{-1}$) the $g^{(2)}$ correlation function can be approximated under the integral \eqref{eq:ffunction} by a constant function equal to 1.  In this limit, the energy functional can be approximated as
\begin{equation}
    \varepsilon(\rho) = \frac{\hbar^2 \pi^2}{6m} \rho^2 - \frac{\lambda}{2} \rho,
\end{equation}
and it is clear that there exists a minimum, namely
\begin{equation}\label{eq:minimum}
    \rho_0=\frac{3m \lambda}{2 \hbar^2 \pi^2}.
\end{equation}
There is also a minimum for $\rho=0$, but the energy corresponding to finite density minimum~\eqref{eq:minimum} is lower $\varepsilon(\rho_0)=-3 m \lambda^2/(8 \hbar^2 \pi^2)<0$.
Hence, in this case, the atoms at the ground state may form a droplet, with the bulk density equal to  $\rho_0$.

Now we wish to go beyond that limit and look for local minima of the energy functional for finite $\kappa$. The condition for extrema of $\varepsilon(\rho)$ 
\begin{equation}
    \varepsilon'(\rho)=\frac{\hbar^2 \pi^2}{3m} \rho -\frac{\lambda}{2} f_\mathcal{V}( \rho \sigma) -\frac{\lambda}{2} \rho \sigma f_\mathcal{V}'(\rho \sigma)=0
\end{equation}
can be rewritten as
\begin{equation}
\label{eq:1stder}
    \alpha \kappa_0=f_\mathcal{V}(\kappa_0)+\kappa_0 f_\mathcal{V}'(\kappa_0),
\end{equation}
with dimensionless parameter $\alpha=2 \hbar^2 \pi^2 /(3 m \lambda \sigma)$ and $\kappa_0=\sigma \rho_0$ is the extremum. We also need to inspect the sign of the second derivative
\begin{equation}
    \varepsilon''(\rho_0) =\frac{\lambda \sigma}{2} \bigg[ \alpha -2 f_\mathcal{V}'(\kappa_0) -\kappa_0 f_\mathcal{V}''(\kappa_0)\bigg ].
\end{equation}
The positions of extrema and their character depend only on one parameter, $\alpha$. Therefore it makes sense to rewrite the energy functional in the following,  slightly different form:
\begin{equation}
    \varepsilon(\rho)=\frac{\lambda}{4} \rho \Big( \alpha \kappa - 2 f_\mathcal{V}(\kappa) \Big),
\end{equation}
from which we find that the sign of $\varepsilon(\rho)$ depends only on $\alpha$ and $\kappa$ [which is fixed in extremum by $\alpha$ due to constraint \eqref{eq:1stder}]. These observations imply that the phase diagram is controlled solely by $\alpha$. 
\begin{figure}[H]
    \centering
    \includegraphics[scale=0.9]{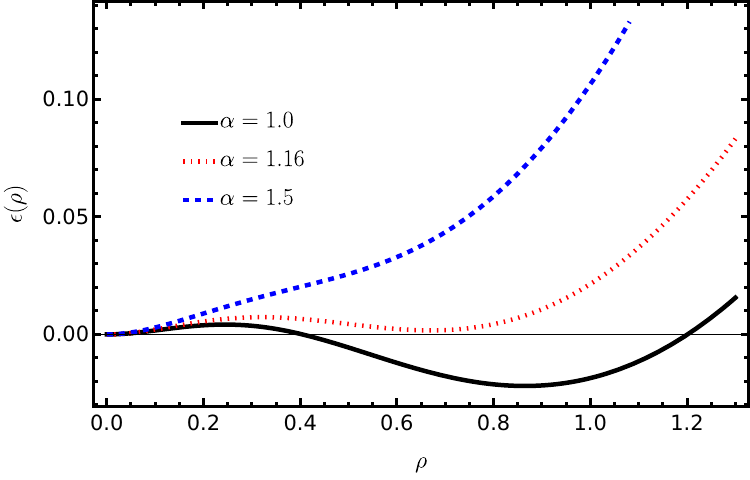}
    \caption{Energy functional in three different regimes. For simplicity we have set $\lambda=1$ and $\sigma=1$. The blue and red curves correspond to gaseous phase as the local minimum is achieved at $\rho=0$. The black curve represents the liquid configuration as the global minimum happens to be non-zero density.}
    \label{fig:2}
\end{figure}
We plot $\varepsilon(\rho)$ for a number of exemplary parameters in Fig. \ref{fig:2}. There are two possible scenarios: either the function has a global minimum for $\rho=0$ (this corresponds to the \textit{gas phase}) or for a finite density (this corresponds to the \textit{liquid phase}). The first scenario can be realized in two ways: $\varepsilon(\rho)$ can have a local minimum for non-zero density (but with positive value of energy, cf.\ the red curve in Fig. \ref{fig:2}) or no finite-density local minimum at all (cf.\ the blue curve on Fig. \ref{fig:2}). For the other scenario (liquid phase), see the black curve in Fig. \ref{fig:2}. 

\begin{figure}[H]
    \centering
    \includegraphics{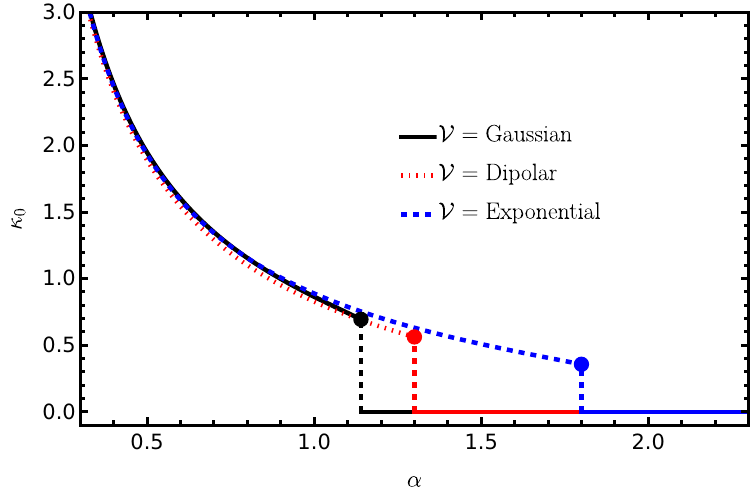}
    \caption{Position of rescaled minimum of energy functional $\kappa_0=\sigma \rho_0$ as a function of parameter $\alpha$. We see that for all the cases there exists a critical point with $\alpha_{\rm c}$ such that for $\alpha >\alpha_{\rm c}$ the parameter $\kappa_0(\alpha)$ becomes equal to zero. Thus for $\alpha< \alpha_{\rm c}$ the system is in liquid phase, whereas for $\alpha>\alpha_{\rm c}$ we enter the gaseous phase. The points have coordinates $(\alpha_{\rm c},\kappa_{0,min})$. We find $\alpha_{\rm liq}^G \approx 1.14$ for Gaussian potential, $\alpha_{\rm liq}^E \approx 1.8$ for exponential and $\alpha_{\rm liq}^D \approx 1.3$ for dipolar. To these points, we may associate minimal values of rescaled densities ($\kappa_0$'s) which are $\kappa_{0,min}^G \approx 0.69$ for Gaussian, $\kappa_{0,min}^E \approx 0.36$ for exponential and $\kappa_{0,  min}^D \approx 0.56$ for dipolar.}
    \label{fig:3}
\end{figure}

Already in Fig.~\ref{fig:2}, we observe that there should be a \textit{gas-liquid transition} somewhere between $\alpha=1.0$ and $\alpha=1.16$. This transition will be visible in the value of density which minimizes the energy functional for given parameters. We thus determine the $\kappa_0$ minimising the energy functional $\varepsilon(\rho)$ as a function of $\alpha$. The results for the three potentials are presented in Fig. \ref{fig:3}. For all potentials we observe a rapid drop of $\kappa_0(\alpha)$ to zero at some critical value $\alpha_{\rm c}$, which depends on $\mathcal{V}$. This behaviour has a simple interpretation: for $\alpha<\alpha_{\rm c}$ (in other words, for sufficiently strong attraction) the system is in liquid phase with non-zero optimal density fixed by the value of $\kappa_0(\alpha)$. On the other hand, for $\alpha>\alpha_{\rm c}$ the system is in gaseous phase as the zero density configuration minimizes the energy. Note that there is a minimal possible value of $\kappa_0$ in the liquid phase, denoted $\kappa_{0, {\rm min}}$.

Fig.~\ref{fig:3} and the discussion above close the analysis of phase diagram in homogeneous system, which turns out to be particularly simple and controlled by a single parameter $\alpha$, for fixed potential $\mathcal{V}$. We have found the liquid phase and in the following sections we will investigate the system in inhomogeneous settings, where the gas tends to form a droplet.

\section{Inhomogeneous systems\label{sec:inhomogeneous}}
In this section we propose an energy functional for inhomogeneous systems. The functional is based on the homogeneous analysis from the previous section. The overall energy has three contributions: kinetic, contact and the one stemming from long-range interactions. Let us start with the last contribution. When the system is inhomogeneous, we can characterise it with space-dependent density field $\rho(x)$ and the long-range contribution to the energy is
\begin{equation}
    E_{\rm LR}[\rho]= \frac{1}{2} \int \dd x \dd x' \, V_\sigma(x-x') g^{(2)}_{TG}(\tilde{\rho};x-x') \rho(x) \rho(x'), \qquad \tilde{\rho}= \rho \left( (x+x')/2\right ).
\end{equation}
Note that this form essentially assumes an approximate ansatz for two-particle correlations:
\begin{equation}
    \langle \hat{n}(x) \hat{n}(x') \rangle = g^{(2)}_{\rm TG}(\tilde{\rho};x-x') \rho(x) \rho(x').
\end{equation}
We proceed with our analysis and introduce new variables $ R=\frac{x+x'}{2}$, $r=x-x',$ getting
\begin{equation}
\begin{aligned}
    E_{\rm LR}[\rho]&= \frac{\lambda}{2 \sigma} \int \dd R \dd r  \, \mathcal{V}(r/\sigma)  g^{(2)}_{TG}(\rho(R);r) \rho(R+r/2) \rho(R-r/2)=
    \\
    & =\frac{\lambda}{2} \int \dd R \dd t \,  \mathcal{V}(t) g^{(2)}_{TG}(\rho(R); t \sigma) \rho(R+t \sigma/2) \rho(R-t \sigma/2),
\end{aligned}
\end{equation}
where we have changed variables as $r=t \sigma$. We look now at expansion of the densities
\begin{equation}
    \rho(R+t \sigma/2)= \rho(R) \left(1+ \frac{\rho'(R)}{\rho(R)}\frac{t \sigma}{2} + \frac{\rho''{R}}{\rho(R)} \left( \frac{t \sigma}{2} \right)^2 + \ldots\right)
\end{equation}
to each term in this expansion we can associate a length scale
\begin{equation}
    l_n(R) =\Bigg|\frac{\rho(R)}{\rho^{(n)}(R)}\Bigg|^{1/n}, 
\end{equation}
related to $n$-th derivative $\rho^{(n)}(R)$ of $\rho(R)$, such that
\begin{equation}
    \rho(R+t \sigma/2)= \rho(R) \left(1+ \text{sgn}(\rho^{(1)}(R))\frac{\sigma}{l_1(R)}\frac{t}{2} + \text{sgn}(\rho^{(2)}(R))\left(\frac{\sigma}{l_2(R)}\right)^2 \left( \frac{t}{2} \right)^2 + \ldots\right).
\end{equation} Our assumptions of slowly varying state is that 
\begin{equation}
    \sigma \ll l_n(R),
\end{equation}
for all $n$ and $R$. Within this expansion, non-locality vanishes in the leading order from our energy functional and in leading order one gets
\begin{equation}
\begin{aligned}
    E_{\rm LR}[\rho]=\frac{\lambda}{2} \int \dd R \,  f_\mathcal{V}\left(\kappa(R)\right) \rho(R)\rho(R), \qquad \kappa(R)=\rho(R) \sigma,
\end{aligned}
\end{equation}
where $f_\mathcal{V}(\kappa)$ was introduced earlier in \eqref{eq:ffunction}. This expression  becomes even simpler if one assumes that $\kappa(R) \gg 1$
for all $R$. Then function $f_\mathcal{V}(\kappa(R)) \to 1$ and contribution from long-range interactions can be written as
\begin{equation}
    E_{\rm LR}[\rho]=\frac{\lambda}{2} \int \dd R \, \rho(R)\rho(R).
\end{equation}
As we will see later, in this limit the system becomes exactly solvable. Let us summarise now the assumptions made in this part. The condition of small gradients $\sigma \ll l_n(R)$ implies slow variation of density in our system and condition $\kappa(R) \gg 1$ -- that the characteristic range of interaction is much bigger than interparticle spacing. These two conditions are independent of each other and set the validity range for our model
\begin{equation}
    \frac{1}{\rho(R)} \ll \sigma \ll l_n(R).
\end{equation}
We proceed with our construction of the energy functional and move to the contribution stemming from the local Fermi pressure. As indicated earlier, we consider the gas in the small gradient limit, hence it is natural to propose the following form of the energy functional for the local kinetic energy (Tonks-Girardeau energy)
\begin{equation}
    E_{\rm TG}[\rho]=\int \dd x  \frac{\hbar^2}{2m} \, \frac{\pi^2}{3} \rho^3(x).
\end{equation}
As the last contribution to the energy we introduce hydrodynamic velocity field of the gas $v(x)$. Kinetic energy of the `envelope' then reads
\begin{equation}
    E_{\rm kin}[\rho,v]= \frac{1}{2} m\int \dd x \rho(x) v(x)^2. 
\end{equation}
The total energy in inhomogeneous system $E=E_{\rm kin}+E_{\rm TG}+ E_{\rm LR}$ is thus a functional of the fields $\rho(x), v(x)$. To make the connection with more standard energy functionals known from cold bosons, such as the Gross-Pitaevskii energy, we introduce a new complex field $\phi(x)$, which is normalised to unity and reads
\begin{equation}
    \phi (x) =\sqrt{\rho(x)/N} e^{i \varphi(x)},
\end{equation}
where $\hbar \partial_x \varphi(x)=m\, v(x)$. Expressing our energy functional with $\phi(x)$ we find
\begin{equation}\label{eq:enefunctional}
    E[\phi] = N\int \dd x \frac{\hbar^2}{2m} \left|\frac{\partial \phi}{\partial x}\right|^2 + N^3\int \dd x \frac{\hbar^2}{2m} \frac{\pi^2}{3} \left|\phi(x)\right|^6 + N^2\frac{\lambda}{2} \int \dd x \left|\phi(x)\right|^4.
\end{equation}
In this step we have neglected the so-called quantum pressure term, which is higher order in gradients of $\rho(x)$~\cite{Damski2004}.
Before we move on, let us mention that one can give up the condition $\sigma \ll l_n(R)$ (but still maintaining $\frac{1}{\rho(R)} \ll \sigma$) which results in energy functional with explicit non-local interaction term
\begin{equation}\label{eq:enefunctional_nonlocal}
    E_{\rm n.local}[\phi] = N\int \dd x \frac{\hbar^2}{2m} \left|\frac{\partial \phi}{\partial x}\right|^2 + N^3\int \dd x \frac{\hbar^2}{2m} \frac{\pi^2}{3} \left|\phi(x)\right|^6 + \frac{N^2}{2} \int \dd x \dd y V_\sigma (x-y) \left|\phi(x)\right|^2 \left|\phi(y)\right|^2.
\end{equation}
Throughout the paper, will focus on analytically solvable case with energy functional \eqref{eq:enefunctional}. However sometimes we will compare them with results obtained with \eqref{eq:enefunctional_nonlocal}.

We go back to \eqref{eq:enefunctional} and minimise the energy functional $\delta E/ \delta\phi^*=0$ under the condition
\begin{equation}
    \int \dd x |\phi(x)|^2=1,
\end{equation}
which amounts to introducing a Lagrange multiplier $\mu$, which has an interpretation of chemical potential in the system. The stationary state equation is 
\begin{equation}
 \label{eq:LLGP_analytic}
      \mu \phi=-\frac{1}{2} \frac{\partial^2 \phi}{\partial x^2}+\frac{\pi^2N^2}{2}|\phi|^4\phi-\lambda N|\phi|^2\phi,
\end{equation}
which was studied also in  \cite{baizakov_solitons_2009}. 
We use box-like units -- we assume an arbitrary length $L$ as the unit of length, and then $mL^2/\hbar$ and $\hbar^2/(mL^2) $ as the
units of time, and energy, respectively.
Note, that the resulting equation contains two competing nonlinear terms, that are different than the nonlinearities appearing for weak interactions only \cite{lima_beyond_2012}.  In what follows, we will study elementary excitations (Sec.~\ref{sec:exc}) and dynamics of a perturbed droplet (Sec.~\ref{sec:nonlin}), using as a starting point a time-dependent version of Eq.~\eqref{eq:LLGP_analytic}, that reads:
 \begin{align}
 \label{eq:LLGPdyn}
    \frac{\partial \phi}{\partial t}=-\frac{1}{2} \frac{\partial^2 \phi}{\partial x^2}+\frac{\pi^2N^2}{2}|\phi|^4\phi-\lambda N|\phi|^2\phi.
 \end{align}
\begin{figure}[h]
    \centering
\includegraphics[width=0.9\textwidth]{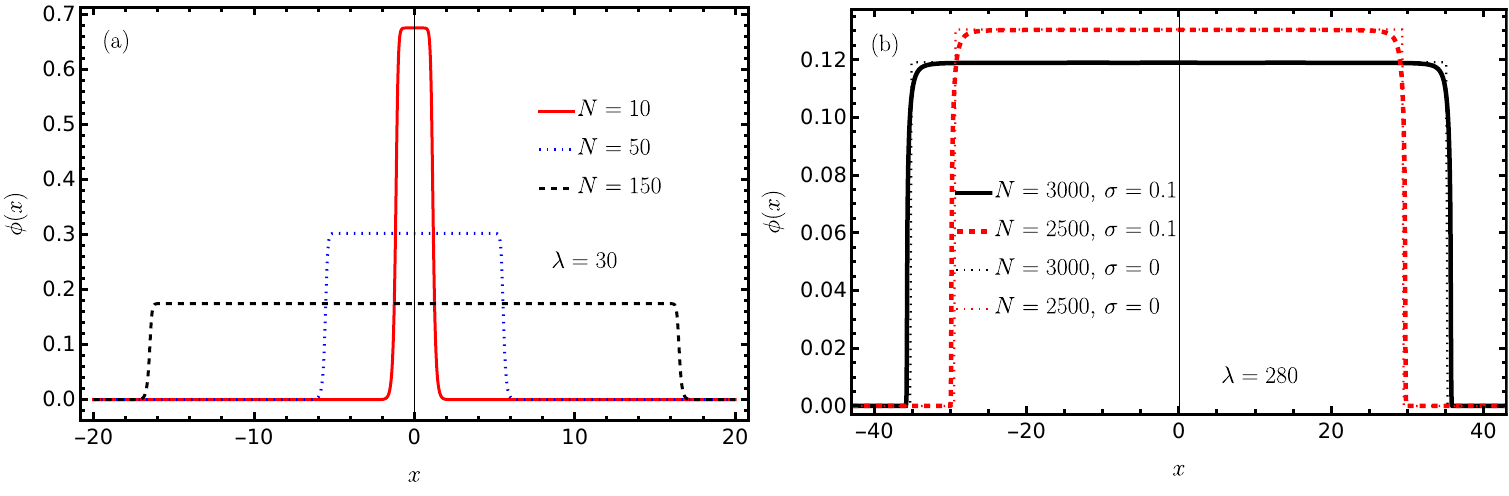}
    \caption{Panel (a): analytical solutions \eqref{eq:LLGP_analytic_sol}  for different particle numbers $N$ and fixed $\lambda=30$. The droplet becomes wider with increasing $N$. Moreover, the density profiles become dominated by the bulk, flat-top part and approach a rectangular shape. Panel (b): Numerical solutions for dipolar potential with finite $\sigma$ obtained from energy functional \eqref{eq:enefunctional_nonlocal}.}
    \label{fig:4}
\end{figure}
The ground state of Eq.~\eqref{eq:energy-functional-TG} has been derived in the context of the nonlinear optics \cite{Pushkarov1979Nov,baizakov_solitons_2009}, and it reads
\begin{equation}
\label{eq:LLGP_analytic_sol}
    \phi_0 (x)=\sqrt{\frac{\sqrt{3}\lambda}{2\pi \eta}} \frac{\tanh (\eta)}{\sqrt{1+\text{sech} (\eta) \,\cosh (x/a)}},
\end{equation}
where 
\begin{equation}
    \eta =\frac{\pi N}{\sqrt{3}}, \qquad a = \frac{\pi}{\sqrt{3} \lambda\tanh (\eta)}.
\end{equation}
Plots of the solution \eqref{eq:LLGP_analytic_sol} for some exemplary parameters are presented in Fig.~\ref{fig:4}.
We also derived the total energy \eqref{eq:ene_analytic} of the ground state solution \eqref{eq:LLGP_analytic_sol}, in terms of its three contributions,  corresponding to the kinetic energy of the envelope $E_{\rm kin}$, the Tonks-Girardeau energy $E_{\rm TG}$ and the long-range $E_{\rm LR}$ energy. The explicit formulas are 
\begin{equation}
\label{eq:ene1_analytic}
    E_{\rm kin}=\frac{3\sqrt{3}}{16 \pi^3}\lambda^2 \tanh \eta \Bigg( -\frac{\eta}{\cosh \eta \sinh \eta} +1 \Bigg),
\end{equation}
\begin{equation}
    E_{\rm TG}=\frac{3\sqrt{3}}{16\pi^3} \lambda^2 \tanh \eta \Bigg( (2+\text{sech}^2 \eta) \eta \coth \eta -3 \Bigg),
\end{equation}
\begin{equation}
    E_{\rm LR}=-\frac{12\sqrt{3}}{16\pi^3} \lambda^2\tanh \eta \Bigg( \eta \coth \eta  -1 \Bigg).
\end{equation}
In the limit of large number of atoms one obtains  a simple expression for the total energy (see Fig.~\ref{fig:5} for comparison).
\begin{equation}
    E= E_{\rm kin} + E_{\rm TG} + E_{\rm LR} \approx -\frac{3\lambda^2}{8 \pi^2}N +\frac{3\sqrt{3}\lambda^2}{8 \pi^3}.
\label{eq:ene_analytic}
\end{equation}
Note that the $N$-independent term in the context of one-dimensional droplets can be interpreted as surface tension energy. It contains contribution from all types of energies. Its precise value may be useful for the study of phenomena related to fragmentation of droplets~\cite{de_palo_formation_2022}, as it gives the amount of energy required to split a droplet into two smaller ones.
\begin{figure}[h]
    \centering
    \includegraphics[width=0.8\textwidth]{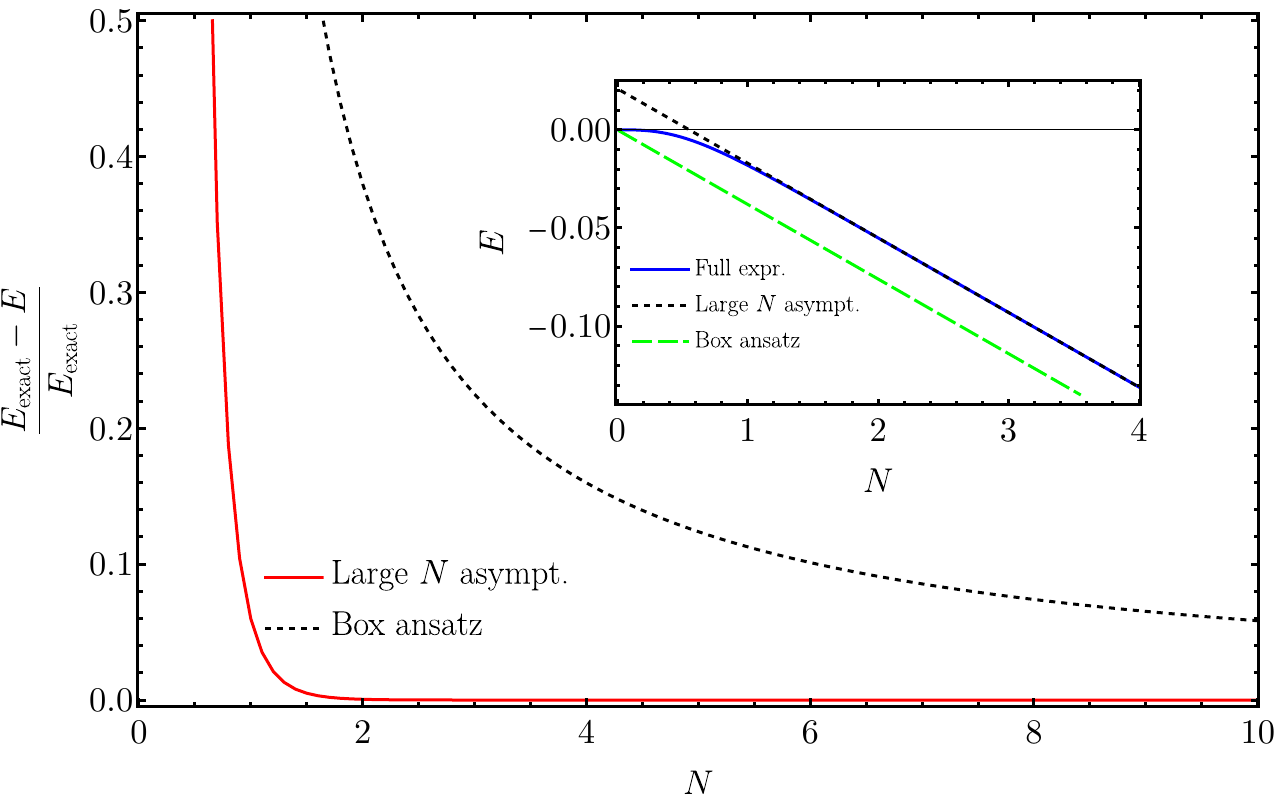}
    \caption{Energy differences between exact energy $E$, its large $N$ asymptotics [the right hand side of Eq.~\eqref{eq:ene_analytic}] and the energy \eqref{eq:exc_rect} of the rectangular ansatz approximation with fixed $\lambda=1$ and as function of $N$. The differences are negligible for large enough $N$. Inset: Comparison between the exact energy $E$ together with large $N$ asymptotics and rectangular ansatz formula \eqref{eq:muEneBox_analytic}.}
    \label{fig:5}
\end{figure}
Looking at Fig.~\ref{fig:4}, one can notice that when $N$ becomes larger and larger, the solution \eqref{eq:LLGP_analytic_sol} starts to resemble a rectangle. This motivates us to introduce the following,  very simple ansatz
\begin{equation}
\label{eq:rect_ansatz}
    \phi_W(x) =\frac{1}{\sqrt{W}} \text{rect}(x/W),
\end{equation}
where the width $W$ is the only parameter of the ansatz. 
We plug our ansatz to the energy functional Eq.~\eqref{eq:enefunctional} and minimise energy with respect to $W$. Additionally, we neglect term involving $\frac{\partial \phi}{\partial x}$, in the same spirit as neglecting the kinetic energy in the Thomas-Fermi approximation. Namely, in the number of parameters we checked that the kinetic energy is much smaller  than interaction energy,  stemming from the other two terms. This can be also confirmed by the analytical solutions. The asymptotic form of the kinetic energy \eqref{eq:ene1_analytic} becomes $N$-independent there, and thus its contribution to the whole energy \eqref{eq:ene_analytic} is negligible for large $N$. 
Under this assumption, we look for the width $W$ minimising energy
\begin{equation}
    \frac{\dd E}{\dd W}=-\frac{N^3}{W^3}\frac{\pi^2}{3}+\lambda\frac{N^2}{2W^2}=0,
\end{equation}
from which we find
\begin{equation}
     W=\frac{2 \pi^2 N}{3\lambda}.
 \end{equation}
Additionally, we may calculate the corresponding chemical potential and energy in the rectangular approximation
\begin{equation}
\label{eq:muEneBox_analytic}
    \mu_W =-\frac{3\lambda^2}{8 \pi^2}, \qquad E_W=-\frac{3\lambda^2N}{8 \pi^2}.
\end{equation}
We can directly compare rectangular ansatz energy to the exact expressions. From comparison of Eqs.~\eqref{eq:ene_analytic} and \eqref{eq:muEneBox_analytic}, we see that rectangular ansatz gives correct value of the term proportional to $N$ but neglects $N$-independent constant corresponding to surface tension energy. This is in agreement with our initial assumption of large $N$. In Fig.~\ref{fig:5}, we summarise the quality of the large $N$ approach and the rectangular ansatz by showing the relative error in the energy computed using these approaches compared to the exact formula. For as little as $N=30$ atoms all approaches give practically the same energy.

The rectangular ansatz leads to simple analytic formulas characterizing parameters of a droplet. In the remaining part of the paper, we use it further to derive and understand better the elementary excitations in the system.
\section{Elementary excitations}\label{sec:exc}
The main goal of this paper is to study excitations of 1D droplets. To access elementary excitations and energies, we start with a time-dependent equation \eqref{eq:LLGPdyn} and then linearise it, as described below.
 
The linearisation procedure, on a formal level, is equivalent to the standard Bogoliubov-de~Gennes (BdG) framework applied to nonlinear Schr\"odinger-like equations \cite{kaiser_bose-einstein_2001}. In general, we consider a stationary solution $\phi_0(x)$ to the Eq.~\eqref{eq:LLGPdyn} and wish to characterise a response to some small perturbation of that field.  We consider the following, standard ansatz for the perturbed time-dependent field
\begin{equation}
\label{eq:linansatz}
    \phi(x,t)=\Big(\phi_\text{0}(x)+\delta \phi(x,t)\Big)e^{-i\mu t},
\end{equation}
where $\mu$ is the chemical potential of stationary solution $\phi_0(x)$ and the small dynamical perturbation is assumed in the form \cite{pitaevskii_bose-einstein_2016,kaiser_bose-einstein_2001}
\begin{equation}
\label{eq:deltaansatz}
    \delta \phi(x,t)= u(x)e^{-i\epsilon t}+v^*(x)e^{i\epsilon t},
\end{equation}
where the functions $u(x)$ and $v(x)$ characterise the shape of the perturbation and $\epsilon$ denotes the excitation energy. Derivation of BdG equations in our case amounts to substituting the perturbed state, as given in Eq.~\eqref{eq:linansatz}, to the equation of motion Eq.~\eqref{eq:LLGPdyn} maintaining the $0$-th and $1$-st order powers of $\delta\phi$ and  $\delta\phi^*$. We get
\begin{subequations}
\begin{align}
    &\epsilon u(x)= \Big( -\frac{1}{2}\partial_x^2 -\mu +\frac{3}{2} \pi^2N^2 \phi_0^4(x) - 2 \lambda N \phi_0^2(x) \Big) u(x)+\Big(\pi^2N^2 \phi_0^4(x) - \lambda N \phi_0^2(x)\Big) v(x),
    \label{eq:bdguv1}
\\
    -&\epsilon v(x)= \Big( -\frac{1}{2}\partial_x^2 -\mu +\frac{3}{2} \pi^2N^2 \phi_0^4(x) - 2 \lambda N \phi_0^2(x) \Big) v(x)+\Big(\pi^2N^2 \phi_0^4(x) - \lambda N \phi_0^2(x)\Big) u(x).
    \label{eq:bdguv2}
\end{align}
\end{subequations}
Following~\cite{Ronen2006}, we introduce new variables $r(x):=u(x)+v(x)$ and $s(x):=v(x)-u(x)$ transforming our equations to
\begin{equation}
    \begin{pmatrix}
    0 & \hat{\mathcal{A}} \\
    \hat{\mathcal{B}} & 0
    \end{pmatrix}\begin{pmatrix}
  r\\
  s
  \end{pmatrix}=\epsilon \begin{pmatrix}
  r\\
  s
  \end{pmatrix},
  \label{eq:rsrelation}
\end{equation}
with 
\begin{subequations}
\begin{align}
    &\hat{\mathcal{A}}:=-\frac{1}{2}\partial_x^2 - \mu + \frac{1}{2}\pi^2N^2 \phi_0^4(x)-\lambda N \phi_0^2(x),
\\
    &\hat{\mathcal{B}}:=-\frac{1}{2}\partial_x^2 - \mu + \frac{5}{2}\pi^2N^2 \phi_0^4(x)-3\lambda N \phi_0^2(x).
\end{align}
\end{subequations}

We further simplify our problem by acting with the matrix above once again on the Eq. \eqref{eq:rsrelation}. We get
\begin{subequations}
\begin{align}
    &\hat{\mathcal{A}}\hat{\mathcal{B}}r(x)=\epsilon^2r(x),
    \label{eq:bdgfinal1}
\\
    &\hat{\mathcal{B}} \hat{\mathcal{A}}s(x)=\epsilon^2s(x).
    \label{eq:bdgfinal2}
\end{align}
\end{subequations}
Note that it is sufficient to solve the eigenproblem for only one equation \eqref{eq:bdgfinal1} or \eqref{eq:bdgfinal2}. The remaining eigenvectors (either $s(x)$ or $r(x)$) can be found from the relation \eqref{eq:rsrelation} between 
$r(x)$ and $s(x)$. The normalisation of $u(x)$ and $v(x)$ is given by the standard condition
\begin{equation}
\label{eq:uvnorm}
    \int {\rm d}x \Big(|u(x)|^2-|v(x)|^2\Big)=\frac{1}{N},
\end{equation}
that for modes with non-zero energy may be expressed with a condition for $r(x)$ function as $\epsilon^{-1}\int {\rm d}x  \,r(x) \hat{\mathcal{B}} r(x) =\frac{1}{N}$.

Note that using Eq.~\eqref{eq:linansatz}, we may write down the approximate time evolution of a density profile
\begin{equation}
\label{eq:densityPert}
    \rho(x,t)=|\phi(x,t)|^2\approx|\phi_0(x)|^2+2\phi_0(x)\Big(u(x)+v(x)\Big)\cos(\epsilon t).
\end{equation}
Therefore, function $r(x)=u(x)+v(x)$ is directly related to the shape of the perturbation of the density profile.

The BdG equations \eqref{eq:bdgfinal1} and \eqref{eq:bdgfinal2} are solved in the momentum space. We discretise the space and work on finite matrices. We find that the low-lying energies converge when one takes sufficiently large number of the numerical grid points.

\begin{figure}[h]
    \centering
    \includegraphics[width=0.85\textwidth]{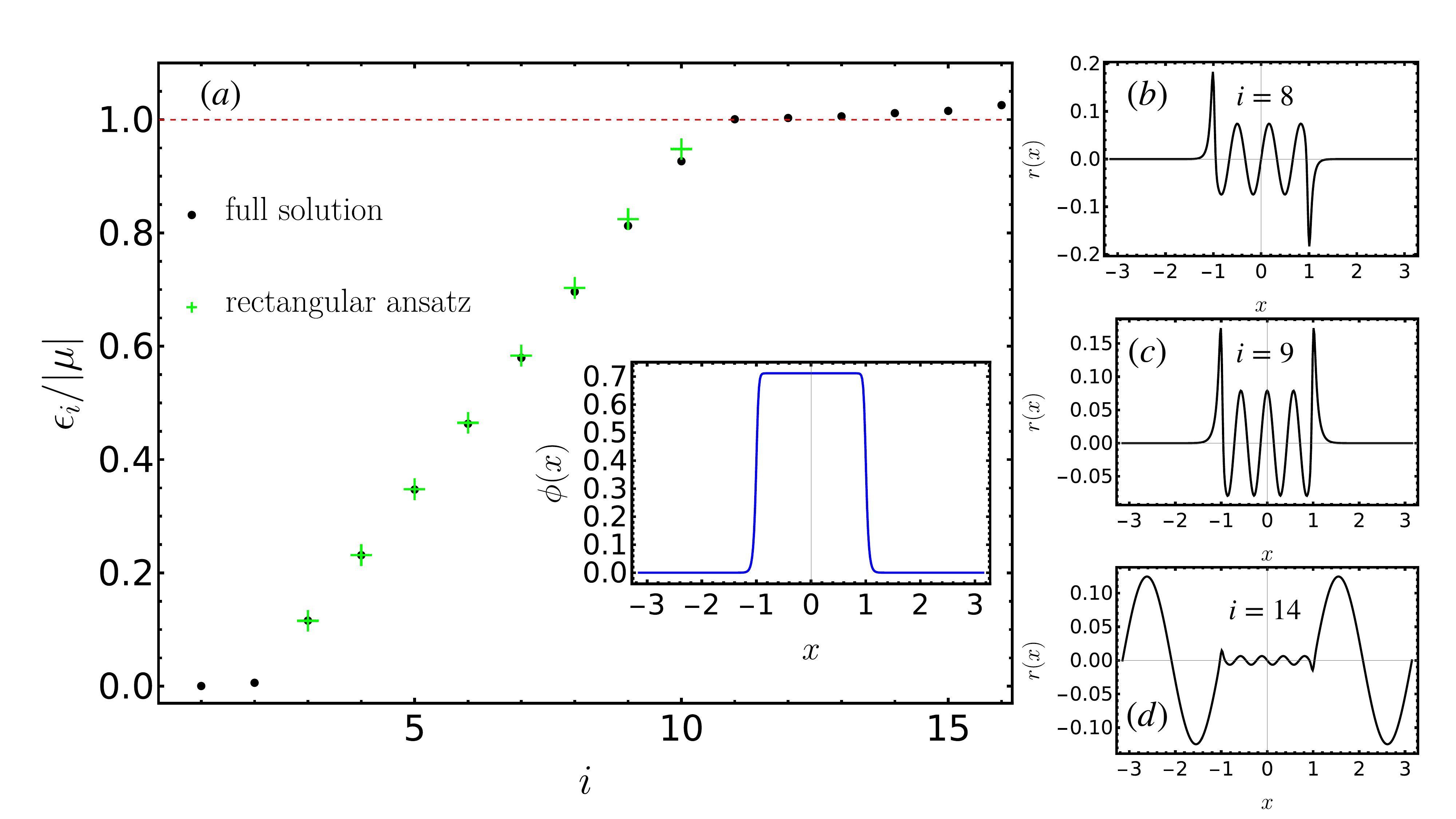}
    \caption{Panel (a): excitation energies found from the solution of BdG equations for $N=30$ and $\lambda=100$. There are two zero-energy modes and eight bound modes with energies below chemical potential. The rest can be classified as scattering modes. In the inset we present density profile of the stationary solution. Additionally, we plot excitation energies found in the rectangular ansatz that match well with the energies of bound modes. Panels (b) and (c): shapes of exemplary bound modes (antisymmetric and symmetric, respectively). Panel (d): an example of a scattering mode. Note the non-zero probability of finding particle outside the droplet in mode (d). }
    \label{fig:6}
\end{figure}

We illustrate our result for the excitations  in Fig.~\ref{fig:6}. In Fig.~\ref{fig:6}(a) we see several lowest-lying excitation energies for an exemplary quantum droplet, whose density profile is presented in the inset of that figure. We have two zero-energy modes related to the breaking of the translational and phase symmetries. Next, we have a couple of modes that we call `bound' as their energy is below  the absolute value of chemical potential and their profiles decay outside of the droplet. These modes are either symmetric or antisymmetric functions of the position [see Figs.~\ref{fig:6}(b,c)]. Importantly, apart from a small region near the edges of the droplet, the modes strongly resemble consecutive standing waves of the infinite box potential with the width given by $W$, implying that the edges of droplets effectively impose open boundary conditions. Above the absolute value of chemical potential, we have scattering modes, with a non-zero probability of finding a particle outside the quantum droplet [cf. Fig.~\ref{fig:6}(d)]. The discrete character of the spectrum of these modes is inherited from the imposed periodic boundary conditions. In general, the scattering modes have a continuous spectrum. We checked that by accessing different momenta via varying the size of the periodic box.

 \begin{figure}[h!]
     \centering
     \includegraphics[width=0.65\textwidth]{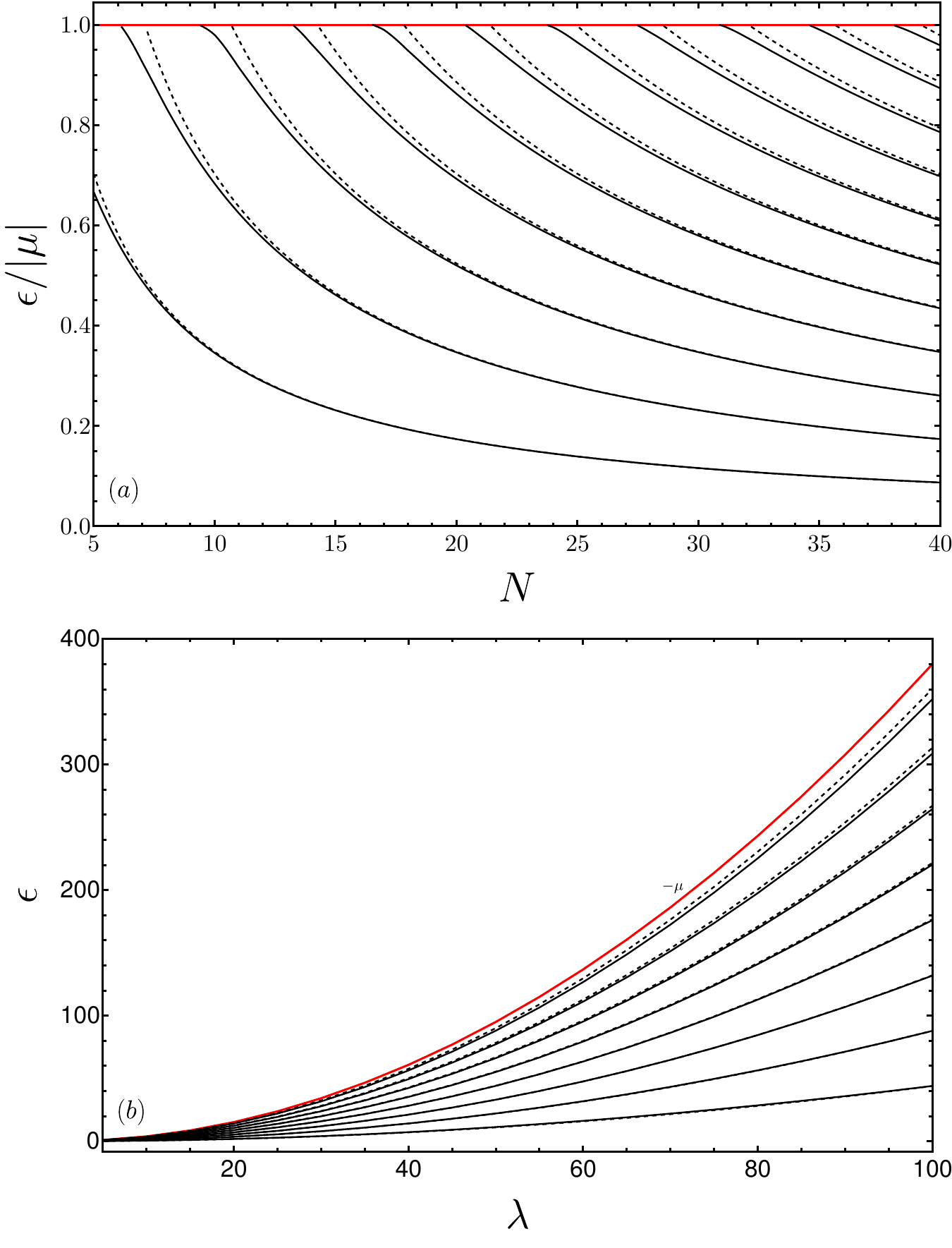}
     \caption{Panel (a): Excitation spectrum of bound modes for a droplet with $\lambda=5$ as a function of particle number $N$. We see that as $N$ grows, the number of bound modes increases. Lines correspond to $m=1$ for the lowest one up to $m=11$ for the uppermost one. Panel (b): Excitation spectrum of bound modes for $N=30$ particles as a function of $\lambda$. The energies are growing functions of $\lambda$. The number of the bound modes does not depend on $\lambda$. In both panels dashed lines mark the results obtained within the framework of rectangular ansatz, Eq.~\eqref{eq:excbox}. Curves correspond to $m=1$ for the lowest one up to $m=8$.} 
     \label{fig:7}
 \end{figure}
 
 The structure of elementary modes turns out to be generic. In the Fig.~\ref{fig:7} we present how excitation spectrum of the bound modes changes when system parameters $N$ and $\lambda$ are varied. We observe that the number of excitation modes grows with $N$ [see Fig.~\ref{fig:7}(a)], but is independent on $\lambda$ [see~Fig.~\ref{fig:7}(b)]. Interestingly, these observations can be deduced from an approximate simple analytical approach, based on the rectangular ansatz introduced earlier.

 \subsection*{Rectangular ansatz approach to the excitations}
 In Sec. \ref{sec:inhomogeneous} we have shown that width, energy and density of a droplet may be approximated by the results based on a simple rectangular ansatz \eqref{eq:rect_ansatz}. Here, we will use  the rectangular ansatz for studying the elementary excitations.  We assume that excitations exist only within the mean-field potential created by atoms forming a droplet. Mathematically the problem is close to the eigenvalue problem for a particle in the box with open boundary conditions.
 We utilise this approach further and solve BdG equations taking $\phi_0(x)=\phi_W(x)$. In this case, BdG equations may be written as
 \begin{subequations}
\begin{align}
    &\epsilon \, u(x)= \Big( -\frac{1}{2}\partial_x^2 +\pi^2 (N/W)^2-\lambda N/W\Big) u(x)+\Big(\pi^2(N/W)^2-\lambda N/W\Big) v(x),
\\
    -&\epsilon\, v(x)=\Big( -\frac{1}{2}\partial_x^2 +\pi^2 (N/W)^2-\lambda N/W \Big) v(x)+\Big(\pi^2(N/W)^2-\lambda N/W\Big) u(x).
\end{align}
\end{subequations}
We are looking for solutions in the form of standing waves $u(x)=u \cos (px)$ and $v(x)=v\cos (px)$ with $p=\frac{\pi}{W}, \frac{2\pi}{W},\ldots$. This choice ensures
that the boundary conditions are satisfied, i.e. the $u(x)$ and $v(x)$ modes vanish at the boundary of the droplet. These equations can be readily solved, separately for every $p$ giving
\begin{equation}
\label{eq:exc_rect}
    \epsilon(p)=\sqrt{\Big((\pi N/W)^2-\lambda N/W\Big)p^2+\Big(\frac{p^2}{2}\Big)^2}.
\end{equation}
Using the formula for the droplet width derived in the rectangular ansatz in the previous Section, $W=\frac{2 \pi^2 N}{3\lambda}$ we obtain explicit excitation energy of $m$-th mode with momentum $p=\frac{m \pi}{W}$
 \begin{equation}
    \epsilon_m=\frac{m}{N} \Big( \frac{3}{2 \pi} \Big)^2\sqrt{\frac{1}{3}+\frac{m^2}{4N^2} }\, \lambda^2.
\label{eq:excbox}
\end{equation}
Hence, excitation energies for fixed $N$ grow quadratically with $\lambda$ [cf. Fig.~\ref{fig:7}(b)]. For large $N$ and fixed $\lambda$, energies decay 
as $\propto N^{-1}$ [cf. Fig.~\ref{fig:7}(a)] and are proportional to $m$, resembling phononic dispersion relation.

In principle, the spectrum given in Eq.~\eqref{eq:excbox} consists of infinitely many eigenenergies. We phenomenologically cut it on the level $\epsilon_{m_\text{max}}=-\mu_W = \frac{3\lambda^2}{8 \pi^2}$. This gives  condition for the maximal number of modes inside a droplet $m_\text{max}$:
\begin{equation}
    m_\text{max}=\sqrt{\frac{1}{3}(\sqrt{5}-2)}N\approx0.28 N.
\end{equation}
The number of  bound modes grows linearly with the particle number $N$ and it is independent from $\lambda$ in accordance with Fig.~\ref{fig:7}(b).

Results from rectangular ansatz approach are compared to the full solution of the BdG equations (solving numerically Eq.~\eqref{eq:bdgfinal1} and\eqref{eq:bdgfinal2}) in Figs.~\ref{fig:7}(a,b) and Fig.~\ref{fig:6}(a). We see a good agreement, especially for the lowest-lying excitations with the eigenenergies far from $-\mu$. The framework built on the rectangular ansatz allows for particularly simple analytical description of droplet excitations that capture the most important features. In particular, excitation modes may be approximately seen as a cosine standing waves of a box with a width given by $W$ and the associated energy of $m$-th mode is determined from \eqref{eq:excbox}.

At this point, it is worth to address two issues. First, analogous excitations were already studied in the droplets in a weakly interacting Bose-Bose mixtures~\cite{tylutki_collective_2020}, that results from the Gross-Pitaevskii equations extended by the LHY terms~\cite{petrov_ultradilute_2016}. That results are qualitatively similar to our findings. In both setups, droplets have bound modes displaying phononic dispersion relation and scattering modes. Moreover, the number of bound modes is finite and scales proportionally to $N$ for large droplets. To understand these analogies, one can note that both equations are similar in the sense that they are local equations with competing nonlinearities. In fact, the only difference lies in the different power-law dependence of the terms corresponding to the repulsion and attraction in the system. 

\section{Nonlinear response to a perturbation}\label{sec:nonlin}

In this section, we will demonstrate the quality of our predictions from the previous sections by numerically studying the dynamics beyond its linearised version. 
As initial state, we take a droplet perturbed with the elementary excitations:
\begin{equation}
    \phi(x;t=0) \propto \phi_0(x) \left( 1+ \delta (\,  u_m(x) + v_m^*(x))\right),
    \label{eq:perturbed-droplet}
\end{equation}
where the small parameter $\delta$ is a perturbation strength and $u_m$ and $v_m$ are (particular) solutions of the BdG equations \eqref{eq:bdguv1} and \eqref{eq:bdguv2}. 
We study numerically the evolution of a system initiated in the state \eqref{eq:perturbed-droplet} using the dynamical equation  \eqref{eq:LLGPdyn} corresponding to the limit $\kappa\to \infty$.
We are interested in the full dynamics of the phonon-like exctiations (low $m$), and the fate of modes with the energy exceeding the threshold $-\mu$.

\begin{figure}[ht!]
    \centering
    \includegraphics[width=0.75\textwidth]{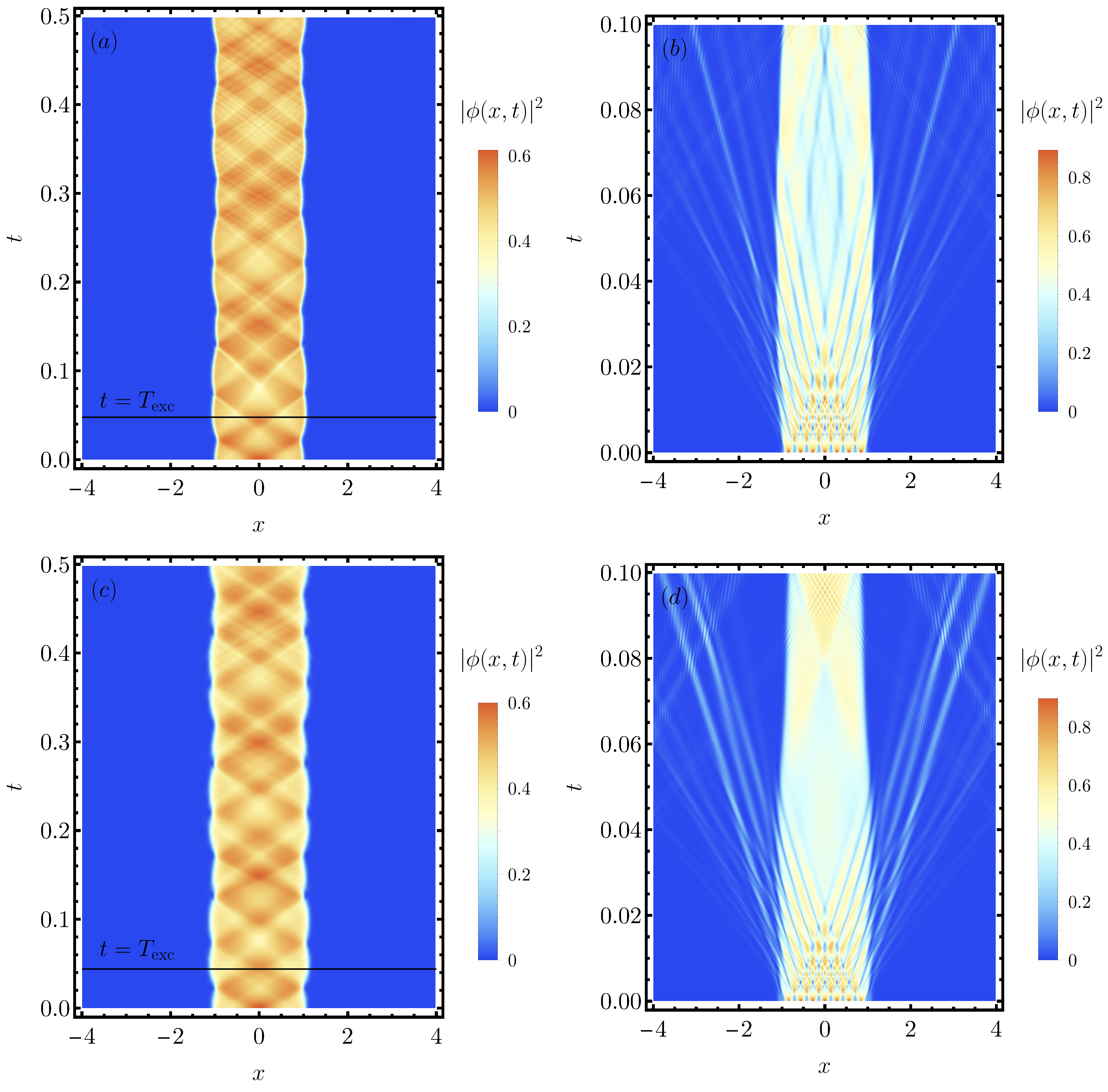}
    \caption{Panels (a,b): dynamics of the droplet with $N=30$, $\lambda=100$  perturbed with excitations with $m=3$ (a) and $m=14$ (b). In the (a) panel we mark with a black solid line the period $T_\text{exc}$ related to the excitation energy. It matches the instant of time at which the system revives to its initial state. In the (b) panel the excitation energy exceeds the chemical potential, and we see the emission of particles in the course of time evolution. Panels (c), (d): similar dynamics but for dipolar interactions with $\sigma=0.05$ and the same $\lambda$ obtained using energy functional \eqref{eq:enefunctional_nonlocal}.}
    \label{fig:fig8}
\end{figure}

The results are presented in Fig. 
\ref{fig:fig8}. We consider there dynamics of a droplet obtained for the same parameters as in Fig. \ref{fig:6}, namely $N=30$ and $\lambda=100$. In Fig.~\ref{fig:fig8}(a) we study the dynamics of a bound mode with $m=3$ with the excitation energy given by \eqref{eq:excbox} and the shape proportional to cosine standing wave $u(x),v(x) \sim \cos \frac{3 \pi x}{W}$.

As the excitation energy is below chemical potential, such a perturbation does not lead to the emission of particles from the droplet. After some time, the nonlinear character of evolution brings additional frequencies to the dynamics. Nevertheless, for short times the evolution is governed by a single frequency, given by the excitation energy. This can be seen from the good agreement between revival time of the excited wave and $T_\text{exc}=2 \pi /\epsilon_3$, cf. black line on Fig. \ref{fig:fig8}(a). The situation is drastically different when we consider a mode with an excitation energy higher than the chemical potential. Droplet for our parameters supports eight bound modes, and perturbing it with a higher energy mode should lead to emission of particles. As we see in Fig.~\ref{fig:fig8}(b), where the dynamics of $m=14$ cosine mode is presented, this is indeed the case. After short time, particles are emitted and the dynamics in the bulk cannot be described as a single wave that propagates inside and is reflected from the edges.

Finally in panels (c) and (d) of Fig.~\ref{fig:fig8} we consider analogous dynamics but obtained with the model \eqref{eq:enefunctional_nonlocal} with long range interactions. We see no qualitative differences as compared to the dynamics in the analytical regime \eqref{eq:enefunctional}.

\section{Summary and outlook}\label{sec:summ}
We studied the general properties and excitation spectrum of a flat-top droplet arising from the interplay between attractive non-local inter-particle forces and strong short-range interactions in 1D.

We showed that such droplets emerge for various non-local interaction potentials. In the hydrodynamic limit of slow density variation with $\kappa = \rho \sigma \gg 1$, where $\sigma$ is the characteristic potential range and $\rho$ is the gas density, the system is well approximated by a local energy functional common to different interaction potentials, with a known analytical solution  for the ground state \cite{baizakov_solitons_2009}. 
We focused on this limit, and studied further the droplet properties and the excitations spectrum. 

In this regime, the excitations spectrum of a droplet arising as a competition between competing forces, turned out to be qualitatively the same as in the  Bose-Bose droplet studied in \cite{tylutki_collective_2020}.
There are two zero-energy modes associated with broken symmetries in the system, and the finite number of bound modes with excitation energies below the absolute value of the chemical potential. The latter modes have roughly phononic dispersion relation and shapes closely resembling standing waves in a box potential with a width determined by the size of the droplet. Finally, the droplet possesses a continuum spectrum of scattering modes characterised by a non-zero probability of finding a particle outside the droplet.
We also drew an analogy between the system with a droplet and a single particle in a potential well with an appropriately defined width and depth, referred to in this paper as the {\it rectangular ansatz}. We used this analogy to derive simple formulas for key quantities characterising the droplet: its total energy, bulk energy, surface energy, width, and excitation spectrum.
We found that  in the strongly interacting regime studied in this paper, the number of phonon-like excitations is approximately $0.28 N$, where $N$ is the number of atoms forming the droplet. We benchmarked our analytical approximations against the numerical simulation of a full (i.e. explicitly non-local) model.

We complemented this analysis with simulations of perturbed droplet dynamics, using time-dependent generalisations of the mean field accounting for the energy obtained within the Lieb-Liniger model \eqref{eq:LLGPdyn}. We confirmed the stability of the droplets against perturbations that excite a bound mode and showed that perturbations with energy above the chemical potential induce particle emission from the droplet.

Mathematically, the droplet in the  regime discussed in this paper shall emerge in a broad class of systems with non-local interaction potentials. However, it remains unclear and shall be investigated, which of the quasi 1D physical system exhibiting non-local interaction can enter in reality the specific regime discussed in the paper --  with strong short-range interaction and $\kappa\gg 1$. The natural candidates are dipolar gases, maybe trapped in an additional optical lattice to change its effective mass, and different implementations of the extended Bose-Hubbard model \cite{Chanda2024May} as indicated in \cite{Morera2021,Marciniak2023}.

\section*{Acknowledgements}
We thank Jan Chwedeńczuk, Maciej Marciniak and Paweł Ziń for valuable discussions.
J.K., M.Ł., and K.P. acknowledge support from the (Polish) National
Science Center Grant No. 2019/34/E/ST2/00289. 
J.K. was supported by the Foundation for Polish Science (FNP) via the START
scholarship.
Center for Theoretical Physics of the Polish Academy of Sciences is a member of the National Laboratory of Atomic, Molecular and Optical Physics (KL FAMO).

\bibliography{bibliography}
\end{document}